\documentclass[preprint, prb, amsmath, aps,showpacs]{revtex4}
\usepackage[dvips]{color}
\newcommand{\bea}{\begin{eqnarray} }
\newcommand{\eea}{\end{eqnarray}}
\newcommand{\bean}{\begin{eqnarray*}}
\newcommand{\eean}{\end{eqnarray*}}
\def\bfg#1{{\mbox{\boldmath $#1$}}}

\def\B{{\bf B}}

\def\J{{\bf J}}

\def\zxi {{\bfg \xi}}

\def\btimes{~{\bf \times}~}
\def\bnabla{{\bf \nabla}}
\def\bcdot{~{\bf \cdot}~}
\def\od#1,#2{\frac{d#1}{d#2}}
\def\odz#1,#2{\frac{d^2#1}{d{#2}^2}}
\def\pd#1,#2{\frac{\partial #1}{\partial #2}}
\def\pdz#1,#2{\frac{\partial^2 #1}{\partial {#2}^2}}
\usepackage{graphicx}
\usepackage{dcolumn}
\usepackage{bm}
\usepackage{epsfig}

\begin{document}

\bibliographystyle{unsrt}
%
%
\title{Diamagnetic drift effects \\on the low-n magnetohydrodynamic modes  
\\at the high mode pedestal with plasma rotation}
%
%
%
\author{L. J. Zheng, M. T. Kotschenreuther, and P. Valanju} 

\affiliation{
Institute for Fusion Studies, 
University of Texas at Austin, Austin, TX 78712}

\date{\today}

\begin{abstract}

The diamagnetic drift  effects on the low-$n$  magnetohydrodynamic instabilities at the high-mode (H-mode) pedestal are investigated in this paper with the inclusion of bootstrap current for equilibrium and rotation effects for stability,  where
$n$ is the toroidal mode number.  The AEGIS  (Adaptive EiGenfunction Independent Solutions) code [L. J. Zheng, M. T. Kotschenreuther, J. Comp. Phys. {\bf 211},  (2006)] is extended to include the diamagnetic drift effects.
This can be viewed as the lowest order approximation of the finite Larmor radius effects  in consideration of 
the pressure gradient steepness at the pedestal.
 The H-mode discharges at  Jointed European Torus (JET) is reconstructed numerically using the VMEC code [P. Hirshman and J. C. Whitson,
Phys. Fluids {\bf 26}, 3553 (1983)],
 with bootstrap current taken into account. Generally speaking, the diamagnetic drift effects are stabilizing. Our results show that the effectiveness of diamagnetic stabilization depends sensitively on the
 safe factor value ($q_s$) at the safety-factor reversal or plateau region. The diamagnetic stabilization are weaker, when
 $q_s$ is larger than an integer; while stronger, when $q_s$ is smaller or less larger than an integer. 
We also find that the diamagnetic drift effects also depend sensitively on the rotation direction. 
The diamagnetic stabilization in  the co-rotation case is stronger than in the counter rotation case with respect to
the ion  diamagnetic drift direction.

\end{abstract}


\maketitle

\section{Introduction}
 
The high mode (H-mode) confinement\cite{hmode} has today been
adopted as a reference for next generation tokamaks, especially
for ITER. However, the H-mode confinement is often
tied with  the damaging edge localized modes (ELMs).\cite{hmode} ELMs can potentially damage divertor plates,
due to  the heat load they cause. This is particularly a challenging issue for a big devices like ITER. Therefore,  the investigation for 
how to mitigate ELMs at the H-mode discharges is important. While various solutions are proposed, 
an interesting solution is to develop the so-called quiescent 
H-mode (QH-mode).\cite{burrell:056121} ELMs are avoided in the QH-modes, due to
 the excitation of the so-called edge harmonic 
oscillations (EHOs) or outer modes (OMs),\cite{burrell:056121,PhysRevLett.104.185003} that pump out plasma energy in
a mild way without exciting the damaging ELMs. 

The current  paper  is aimed at further understanding of ELMs and EHOs 
in the  H-mode or QH-mode discharges. In particular, we investigate the diamagnetic drift effects on low-$n$ ($n=1$, $2$, and $3$) 
magnetohydrodynamic (MHD) instabilities at the H-mode pedestal with the inclusion of bootstrap current for equilibrium and rotation effects for stability,  where $n$ is toroidal mode number. Note that, when bootstrap current is taken into account, a safety-factor reversal or plateau  can be generated 
at the pedestal.\cite{0029-5515-47-9-026} We have shown that the modes of infernal type (or fat interchange modes) \cite{0029-5515-27-9-009,quasii}
can prevail  at the safety-factor reversal or plateau region and found that such a type of 
modes has the typical EHO features at QH-mode discharges.\cite{infernal.pop,infernal.nf} There is a physical ground for us to extend these
investigations to include the diamagnetic drift effects. We note that the ion diamagnetic frequency ($\omega_{*i}$) is directly proportional to 
pressure gradient and inversely proportional to density. This leads the ion diamagnetic frequency $\omega_{*i}$ to become  big and vary 
dramatically at the pedestal, where the infernal modes tend to develop.  Therefore,  the current investigation 
of diamagnetic drift effects is interesting. 

We point out that the peeling or kink/peeling modes were previously 
proposed to explain EHOs or OMs. The diamagnetic stabilization effects on the peeling, ballooning, and peeling-ballooning modes have been studied in Refs.  \onlinecite{zhengpl,hastieflr,nf07,ppcf}. In difference from these investigations we include the bootstrap current effects
on the equilibrium. This inclusion leads the safety factor profile to change\cite{0029-5515-47-9-026} and subsequently
the MHD modes to behave differently. In our earlier works\cite{infernal.pop,infernal.nf} we prove that 
the MHD modes of infernal mode type \cite{0029-5515-27-9-009,quasii} 
can prevail in this case. The diamagnetic stabilization effects
on the infernal modes also are  different, as compared those reviewed  in Ref. \onlinecite{ppcf}.
Based on the ballooning mode investigation,\cite{hastieflr} Reference \onlinecite{ppcf} concludes
that, when there is a radial variation in the diamagnetic frequency $\omega_{*i}$, the diamagnetic stabilization is less effective. Instead,
we find that, when the bootstrap current effect on the equilibrium is taken into account, 
the diamagnetic stabilization can be effective, depending sensitively on the safety factor value
 at  the region where the safety factor is flat or  reversed and also on the toroidal rotation direction.
Besides, we point out that the current investigation is based on the two dimensional free boundary AEGIS  (Adaptive EiGenfunction Independent Solutions) code,\cite{Zheng2006748}
while the previous conclusion about the ineffectiveness of the diamagnetic stabilization is based on
 the conventional one dimensional ballooning representation.\cite{hastieflr,ppcf} As pointed out in
Ref. \onlinecite{freeb} the conventional ballooning mode representation cannot be used
at the plasma edge.  Therefore, the current investigation  extends the existing investigations based on the  peeling-ballooning mode formalism.

This paper is arranged as follows: In Sec.~II the numerical schemes for equilibrium and stability analyses 
are given; In Sec.~III the numerical results are presented;
Conclusions and discussion are given in the last section.

\section{Numerical scheme}

In this section we describe the numerical scheme for equilibrium and stability studies. 
For equilibrium  we focus our investigation on the JET-like QH-mode discharges. The plasma cross section is shown in Fig. 1. The conformal wall is used in our calculations. Furthermore, we  consider only the subsonic rotation case, i.e.,  the rotation frequency is assumed to be much lower than the ion acoustic frequency. In this case the rotational effects of centrifugal force and Coriolis force both on equilibrium
and stability can be neglected. \cite{waelbroeck:601,zheng:1217}
  We then include the rotational effects only through
a Doppler frequency shift in the stability analysis. 

To study the QH mode, we consider the low collisionality regime. 
In this regime the steep pressure gradient at the pedestal region can induce a strong bootstrap current. 
As in Refs. \onlinecite{infernal.pop} and \onlinecite{infernal.nf} 
we use the STELLOPT \cite{spong01} code (part of the VMEC\cite{VMEC} code suite) to compute the collisionless limit of the bootstrap current.\cite{shaing89} For parameters typical of QH mode pedestals with normalized collision frequency $\nu^* = 0.05$ and inverse aspect ratio $r/R = 0.3$, the bootstrap current is approximately 80\%-90\% of the collisionless limit.\cite{landreman12} In addition, numerical calculations find that there are modest modifications due to finite poloidal gyroradius \cite{landreman12} (which have yet to be evaluated for QH modes).  Our equilibrium results are consistent to the previous calculations with other codes as given in Ref. \onlinecite{0029-5515-47-9-026}. 
Taking into account the
bootstrap current in the equilibrium calculation, we found that
a safety-factor ($q$) reversal or plateau   can indeed  appear  in the pedestal region. 
We denote the safety factor value at  the region where the safety factor is flat or  reversed 
as $q_s$. 

Our physics intuition leads to examine the difference between the cases with $q_s$ 
larger and smaller than an integer number. The equilibria are therefore constructed
to have different $q_s$, while minimizing other profile changes. The $q$ profile change is resulted 
from the change of toroidal current. Subsequently, the pressure profile is scaled up or down to keep 
beta normal constant.
We uses five different
equilibria for stability analyses as mostly used in Refs. \onlinecite{infernal.pop} and \onlinecite{infernal.nf} , 
with the safety factor at the plateau, $q_s$, ranging from $4.2$, $4.1$, $4.05$, $4$,  and $3.96$.
The safety factor  and toroidal current profiles are plotted  in Fig. 2;  while
the corresponding pressure profiles are plotted  in Fig. 3. 
We keep the beta normal the 
same in the five equilibria by scaling the overall pressure profile 
appropriately. 

As one can imagine, the appearance of safety factor plateau can minimize the magnetic shear stabilization and cause the  ``fat interchange
modes" (i.e., infernal modes) to develop locally.\cite{infernal.pop,infernal.nf} We consider only the low-$n$ modes:
$n=1$, $2$, and $3$. 
The MHD instabilities in this type of equilibria 
are investigated numerically using the AEGIS code,\cite{Zheng2006748} with both the diamagnetic  and apparent mass effects taken into account.  In including the diamagnetic drift effects we use the frequency modification: $(\omega+n\Omega)^2\to \hat\omega^2\equiv (\omega
+n\Omega)(\omega+n\Omega-\omega_{*i})$, where $\omega$ is the mode frequency and  $\Omega$ is the toroidal rotation frequency.  This modification can be viewed as keeping 
the finite Larmor radius effects in lowest order in consideration of
the  pressure gradient steepness at the pedestal.\cite{zheng2f,zhenggyr}
The adaptive numerical scheme of AEGIS code allows us to study the rotation-induced  continuum damping.\cite{PhysRevLett.95.255003}

The basic MHD equation used for our stability analyses is as follows:
\bean
-\rho_m\hat\omega^2\zxi&=& \delta \J\btimes \B +\J\btimes
\delta \B 
-\bnabla \delta P,
\eean
where   $\zxi$ is the perpendicular 
field line displacement,  $\B$ denotes  the equilibrium magnetic field, 
$\delta \B=\bnabla\btimes\zxi\btimes\B$ is the perturbed magnetic field, $\J$ represents the equilibrium current density, 
$\mu_0\delta \J =\bnabla\btimes 
\delta\B$ is the perturbed current density, 
$\mu_0$ is the magnetic constant,
$P$ represents the equilibrium pressure, $\delta P= -\zxi\bcdot \bnabla P$ is the perturbed pressure of convective part, and the perturbed quantities are tagged
with $\delta$ except $\zxi$. The plasma compressibility effect does not appear in this equation explicitly.  Since
the mode frequencies we are studying are much smaller than the ion acoustic wave frequency,  
the plasma compressibility results only in the so-called apparent mass effect, 
as proved in Ref.  \onlinecite{greene:510}. Therefore, we include the plasma
compressibility effect by regarding $\rho_m$ as the total mass, i.e., the sum of
perpendicular mass and parallel mass (i.e., the apparent mass in the perpendicular momentum equation)
according to Ref.  \onlinecite{greene:510}.

Note that  the EHOs (or OMs) observed experimentally have finite frequencies,
about 10 kHz for $n=1$ modes,\cite{burrell:056121,PhysRevLett.104.185003} which is much larger than
the wall magnetic diffusion time. For modes with such high frequencies 
the wall behaves as a perfect conductor. Therefore, our calculations focus only on the perfectly conducting wall case.  
As in Ref. \onlinecite{infernal.pop} we use the combined methods of the Nyquist diagram and the  analytic continuation of the dispersion relation to determine the unstable roots.

\section{Numerical results}

The diamagnetic drift effects on peeling-ballooning modes have been discussed in Ref. \onlinecite{ppcf}. The discussion
is based on the earlier research about diamagnetic stabilization effects on the ballooning modes in Ref. \onlinecite{hastieflr}. 
When the bootstrap current effects on the equilibrium are taken into account, however, the infernal 
mode  appears and its behavior deviates from that of ballooning modes.  Consequently, one can expect
the diamagnetic drift effects to behave quite differently. To show these different features 
we mainly investigate five equilibria with safety factor at the plateau, $q_s$, ranging from $4.2$, $4.1$, $4.05$, $4$, and $3.96$. From Figs. 2 and 3 one can see that, besides the change in $q_s$, these five equilibria are almost
identical. However, as shown in the following numerical results, the diamagnetic stabilization effects on these equilibria
are vastly different. In our investigation, the rotation frequency profile is assumed to be the same as the pressure profile
and the density profile is assumed to be the same as the temperature one. 

First, we point out that the diamagnetic drift effects are generally stabilizing for infernal modes, as one may expect.
But,  the  effectiveness  of diamagnetic  stabilization depends on  the safety factor value
at the plateau. Figures 5 and 6 show the eigen frequencies and growthrates of $n=1$ modes
versus the wall position respectively for the cases without and with the diamagnetic drift effects. 
The frequencies are normalized by the Alfv\'en frequency at the magnetic axis and the wall position is normalized
by the minor radius in the mid-plane  in this work.  We have assumed
the direction of rotation is the same as the direction of  ion diamagnetic drift motion in most of the investigations
in this work, unless it is indicated otherwise. In these figures the rotation frequency  at the magnetic axis is $\Omega=0.03$ and  the
safe factor value ($q_s$) at the safety-factor reversal or plateau region is used as a parameter. 
The typical real and imaginary eigen functions of $n=1$ modes for the case without the diamagnetic drift effects are shown respectively 
in Figs. 7a and 7b and those for the case with the diamagnetic drift effects are shown respectively 
in Figs. 8a and 8b.  From Figs. 7 and 8 one can see that the $m/n=4/1$ harmonic, which is resonant
at the safety factor plateau,  appears bigger and broad as compared to the usual kink modes.   
Their features as the infernal modes  have been discussed in Refs. \onlinecite{infernal.pop} and \onlinecite{infernal.nf}.
Here, we concentrate on discussing the diamagnetic drift effects. From Figs. 5 and 6  one can see that the growthrates with the diamagnetic drift effects are generally smaller  than those without the diamagnetic drift effects. This shows the general stabilization of the diamagnetic drift effects.
Bigger frequencies  in the case with  the diamagnetic drift effects as compared to the cases without the diamagnetic drift effects are because
 the rotation is assumed to be in the ion diamagnetic drift direction in Fig. 6. 
 From Fig. 6 one can also see that the smaller $q_s$ case has  a larger frequency. This is because, when 
 $q_s$ reduces, the $m=4$ infernal harmonic moves outwardly from the pedestal top and consequently the amplitude 
 of the diamagnetic frequency increases (see the $ \omega_{*i}$ profile in Fig. 4). Notably,
 our results show that the diamagnetic drift effects depend sensitively on the $q_s$ value. The diamagnetic drift effects are weaker, when
 $q_s$ is far larger than an integer (here, it is ``$4$''); while stronger, when $q_s$ is less larger or smaller than an integer.
This can be seen from Fig. 6.  In Fig. 6 the growthrate for $q_s=4.2$ case is one order larger than
the  $q_s=4.1$ case. The modes are stable when $q_s=4.0$ and $3.96$. We also check the further lower $q_s$ case,
for example $q_s=3.92$.  A full diamagnetic stabilization is found as well.
This can be explained by inspecting the profile of ion diamagnetic drift frequency $\omega_{*i}$ in Fig. 4 together with the $m/n=4/1$ infernal harmonic accumulating
point. When $q_s$ is far larger than an integer  ``$4$", the $m=4$ infernal harmonic tends to accumulate at the inner side of  the safety-factor reversal or plateau region. Noting that the ion diamagnetic drift frequency $\omega_{*i}$ is smaller in this region, one can expect a weaker diamagnetic stabilization to result. Whereas, when $q_s$ is less larger than an integer or even smaller than an integer, the accumulation point of the infernal harmonic moves  outwardly relative to  the pedestal top. Noting that the ion diamagnetic drift frequency  $\omega_{*i}$ becomes larger in this region, one can understand why a stronger diamagnetic stabilization should result.

We also explore the diamagnetic drift effects for various rotation speeds. Figures 9 and 10 show the eigen frequencies and growthrates of $n=1$ modes versus the rotation frequency at the magnetic axis respectively for the cases without and with the diamagnetic drift effects.
They are for the equilibrium with $q_s=4.2$. It is interesting to point out that the frequency difference between
the cases with and without the diamagnetic drift effects remains about the same, as the rotation frequency varies. The difference
is about $0.003$. The formula $ (\omega
+n\Omega)(\omega+n\Omega-\omega_{*i})= (\omega
+n\Omega- \omega_{*i}/2 )^2 - \omega_{*i}^2/4$ tells that the $\omega_{*i}$ induced frequency shift is $\omega_{*i}/2$. From the $\omega_{*i} $ profile in Fig. 4 and the eigen mode plots in Fig. 8 one can see that
the location for $\omega_{*i}/2 =0.003$ is around where the $m=4$ Fourier harmonic of infernal mode type locates. This again confirms
the infernal mode feature of the instabilities. 

The rotation direction effects are investigated as well and the results are shown  in Fig. 11. In this figure $q_s=4.1$ and the diamagnetic stabilization effects are
taken into account. Without  the diamagnetic stabilization effects the mode growthrate remains unchanged and only the mode frequency changes signs, as 
 the rotation direction switches. When the diamagnetic stabilization effects are taken into consideration,  Figure 10 shows that 
the frequency difference for two rotation directions is about $0.006$.  Similar to the discussion for Figs. 9 and 10 in the previous paragraph,
this again shows that the  $\omega_{*i}$ value  at where the $m=4$ Fourier harmonic locates
plays a key role --- an infernal mode feature. Figure 11 shows that the stabilization effects for co-rotation case
is much stronger than for  counter-rotation case with respect to
the ion  diamagnetic drift direction. As an infernal mode feature, the mode frequency tends to match the local sum of rotation
and diamagnetic frequencies for $m=4$ infernal  harmonic. However, in the toroidal geometry the sideband effects have to
be considered as well. Note that  rotation and diamagnetic frequencies have radial profiles. 
The combined effects of rotation and diamagnetic frequencies on the sidebands 
between the co-rotation and counter-rotation cases are different. A larger  sum of rotation
and diamagnetic frequencies gives rise to a stronger continuum damping effects.\cite{PhysRevLett.95.255003} This leads
to the counter-rotation case  becomes more unstable than the co-rotation case as shown in Fig. 11.

We have discussed the $n=1$  modes above. Now, we turn to the $n=2$ and $3$ modes. 
Figure 12 gives the dependence of mode frequencies and growthrates for $n=1-3$ modes on the wall position without the
diamagnetic drift effects and Fig. 13 gives the dependence of mode frequencies and growthrates for $n=1$ and $2$ modes on the wall position with the diamagnetic drift effects. Both figures are related to the $q_s=4.2$ equilibrium
and rotation frequency at the magnetic axis is  $\Omega=0.03$. As  pointed out in Ref. \onlinecite{infernal.nf} without the diamagnetic drift effects Fig. 12 shows that the mode frequencies follows the frequency-multiplying rule: $\omega = n\Omega_s$ for $n=1-3$ modes, where $\Omega_s$ is about the rotation frequency at the pedestal top. 
With the diamagnetic drift effects being taken into
account we found that the frequency-multiplying
 rule $\omega = n\Omega_s$ are still roughly kept for $n=1$ and $2$ modes.
The $n=3$ modes, however, do not appear in Fig. 13, since they are stabilized in this equilibrium. Nevertheless, we note that in the nonlinear case the $n=1$ and $2$ modes can couple to give rise to the $n=3$ modes with frequency being
the sum of $n=1$ and $2$ mode frequencies. In view of this the frequency-multiplying rule can still be expected for low-$n$ modes 
 with the diamagnetic drift effects taken into account in the nonlinear description. We also point out that
  in the $n=3$ case (or even $n=2$ case) other kinetic effects may need to be considered, since the diamagnetic drift frequency for $n=3$ is larger.

\section{Conclusions and discussion}

The diamagnetic drift  effects on the low-n magnetohydrodynamic instabilities at the high-mode (H-mode) pedestal are investigated in this paper. In view of  the steep pressure gradient  in the pedestal region, 
the inclusion of  diamagnetic drift effects can be regarded as the inclusion of finite Larmor 
radius effects in the lowest order.\cite{zheng2f,zhenggyr} We focus our investigation on the JET-like QH-mode discharges. Subsonic plasma rotation effects are included in the investigation,
especially the induced continuum damping effect. The differences of current studies from the previous ones based on the ballooning or peeling-ballooning pictures in Refs. \onlinecite{hastieflr}.
and \onlinecite{ppcf} mainly lie in the following aspects: First, we include the bootstrap current effects
on the equilibrium so that a safety factor plateau  is resulted at the pedestal region. Because of that
our focus is shifted to the diamagnetic drift effects on the $m/n=4/1$ infernal harmonic. Second, the current calculation is 
based on the two dimensional free boundary MHD code: AEGIS, while the researches in Ref. \onlinecite{ppcf} is based
on the conventional one dimensional ballooning representation. Because of these differences our research yields several interesting new results, which have not been reported in the previous studies.

First,  we note that the diamagnetic frequency ($\omega_{*i}$) is directly proportional to 
pressure gradient and inversely proportional to density. This leads  the diamagnetic frequency $\omega_{*i}$ to become  big and vary dramatically at
the pedestal, where the infernal modes tend to develop. In view of this fact we find  
that the diamagnetic drift effects depend sensitively on the
 safe factor value ($q_s$) at the safety-factor reversal or plateau region. The diamagnetic stabilization effects are weaker, when
 $q_s$ is larger than an integer; while stronger, when $q_s$ is smaller or less larger than an integer. 
 This is because, when $q_s$ is far larger than an integer, the infernal modes tends to accumulate at the inner side of  the safety-factor reversal or plateau region, where $\omega_{*i}$ is smaller, while  when $q_s$ is smaller or less than an integer the infernal modes tends to move outwardly from the pedestal top,
  where $\omega_{*i}$ becomes larger. A larger  diamagnetic frequency amplitude $\omega_{*i}$  at where the infernal harmonic develops gives
a  stronger stabilization effects. This explains why the $q_s$ value is so critical. 

 We also find that the diamagnetic drift effects depend sensitively on the rotation direction. Counter rotation results
in a weak diamagnetic  stabilization, while co-rotation gives rise to a strong diamagnetic stabilization with respect to
the ion  diamagnetic drift direction. The reason 
is as follows.
The co-rotation results in a larger sum of rotation and
diamagnetic frequencies. Consequently, it leads a stronger continuum damping especially from the sidebands. 

We have studied  the $n=2$ and $3$ modes as well. We find that the $n=3$ modes tends to be stabilized
by the diamagnetic drift effects for the equilibria we considered, because the diamagnetic stabilization effects is proportional to
$\omega_{*i}^2/4$ and $\omega_{*i}$ is directly proportional to the mode number $n$. 
With the diamagnetic drift effects being taken into
account we found that the frequency-multiplying
 rule $\omega = n\Omega_s$ are still roughly kept for $n=1$ and $2$ modes. We discuss that the possible
 nonlinear coupling can lead to the frequency-multiplying rule to hold for $n=3$ modes as well.

The current investigation is focused on the low $n$ modes. In this case the  mode frequency in the rotating frame
for infernal harmonic is still
low, so that the wave-particle resonance effects may be excluded.
In the higher mode number case the wave-particle resonance effects need to be considered.
Due to the dramatic variation of the diamagnetic
frequency at the pedestal, a fully kinetic treatment is strongly preferred for this subject.  
The current investigation may be regarded as an extension of existing researches for diamagnetic stabilization.\cite{hastieflr,ppcf} So far in this field the fully kinetic treatment of pedestal physics
 is still limited to using the fixed boundary ballooning mode formalism. Most of peeling ballooning mode calculations
 are still based on the ideal MHD framework.  Fully kinetic treatment
for free boundary problems is challenging and will be considered in the future.

This research is supported by U. S. Department of Energy, Office of Fusion Energy Science.

\newpage


\newpage

Figure captions:

  Fig.~1:  The cross section of  JET-like configuration. The horizontal coordinate 
$X$ represents the distance from the axisymmetric axis. The vertical coordinate
$Y$ is the height  from the vertical mid-plane. 

  Fig.~2. Safety factor ($q$) and parallel current density ($J$) profiles versus normalized poloidal magnetic flux.  
  The safety factor values at the plateau range from $4.2$, $4.1$, $4.05$, $4$, to $3.96$.

  Fig.~3. Pressure profiles versus normalized poloidal magnetic flux for the cases 
  with the safety factor values at the plateau range from $4.2$, $4.1$, $4.05$, $4$, to $3.96$.
    Pressure is normalized by $ B^2/\mu_0$ at the magnetic axis. 
  
  Fig.~4. The diamagnetic frequency of $n=1$ modes versus  normalized poloidal magnetic flux for the case
  with $q_s=4.1$. Since the pressure profiles for five cases we consider are
  almost identical  as shown in Fig. 3, the diamagnetic frequencies for other
  cases resemble to this figure. 
 
  Fig.~5. Eigen frequencies and growthrates of $n=1$ modes  versus wall position for the cases with $q_s =$ $4.2$, $4.1$, $4.05$, and $4$
  without the diamagnetic drift effects. The four growthrate curves are below the frequency curves.

  Fig.~6. Eigen frequencies and growthrates of $n=1$ modes  versus wall position for the cases with $q_s =$ $4.2$, $4.1$, and $4.05$
  with diamagnetic drift effects. The cases with $q_s=4$, $3.96$, and $3.92$ are stable, when the diamagnetic drift effects are taken into account.
 Note that the growthrate scale for the case with $q_s=4.2$  is 10 times larger as indicated in the figure.

Fig.~7.  The real (a) and imaginary (b)  eigen functions of $n=1$ modes versus normalized poloidal magnetic flux 
  for the case with  $q_p=4.2$, $b=1.5$,  $\Omega=0.03$, and without the diamagnetic drift effects. The $q$ profile
  and the rational surfaces are also given. 

Fig.~8.  The real (a) and imaginary (b)   eigen functions of $n=1$ modes  versus normalized poloidal magnetic flux 
  for the case with  $q_p=4.2$, $b=1.5$,  $\Omega=0.03$, and with the diamagnetic drift effects. The $q$ profile
  and the rational surfaces are also given.

 Fig.~9. Eigen frequencies and growthrates of $n=1$ modes  versus the rotation frequency at the magnetic axis
  for the case with $q_s = 4.2$  without diamagnetic drift effects. 
Wall position is used as the parameter. The three growthrate curves are below the frequency curves.

 Fig.~10. Same as Fig. 9, but the diamagnetic drift effects are taken into account.

 Fig.~11. Eigen frequencies and growthrates of $n=1$ modes  versus wall position for the case with $q_s = 4.2$  with the diamagnetic drift effects being  taken into account.  The results with opposite rotation directions are displayed. The two growthrate curves are below the frequency curves.

 Fig.~12. Eigen frequencies and growthrates  of $n=1-3$ modes  versus wall position 
  for the case with $q_s = 4.2$  and without the diamagnetic drift effects being  taken into account. 
   The three growthrate curves are below the frequency curves.

Fig.~13. Eigen frequencies and growthrates of $n=1$ and $2$ modes  versus wall position 
  for the case with $q_s = 4.2$  and with the diamagnetic drift effects being  taken into account. 
   The two growthrate curves are below the frequency curves.

\newpage

\begin{figure}[htp]
\centering
\includegraphics[width=120mm]{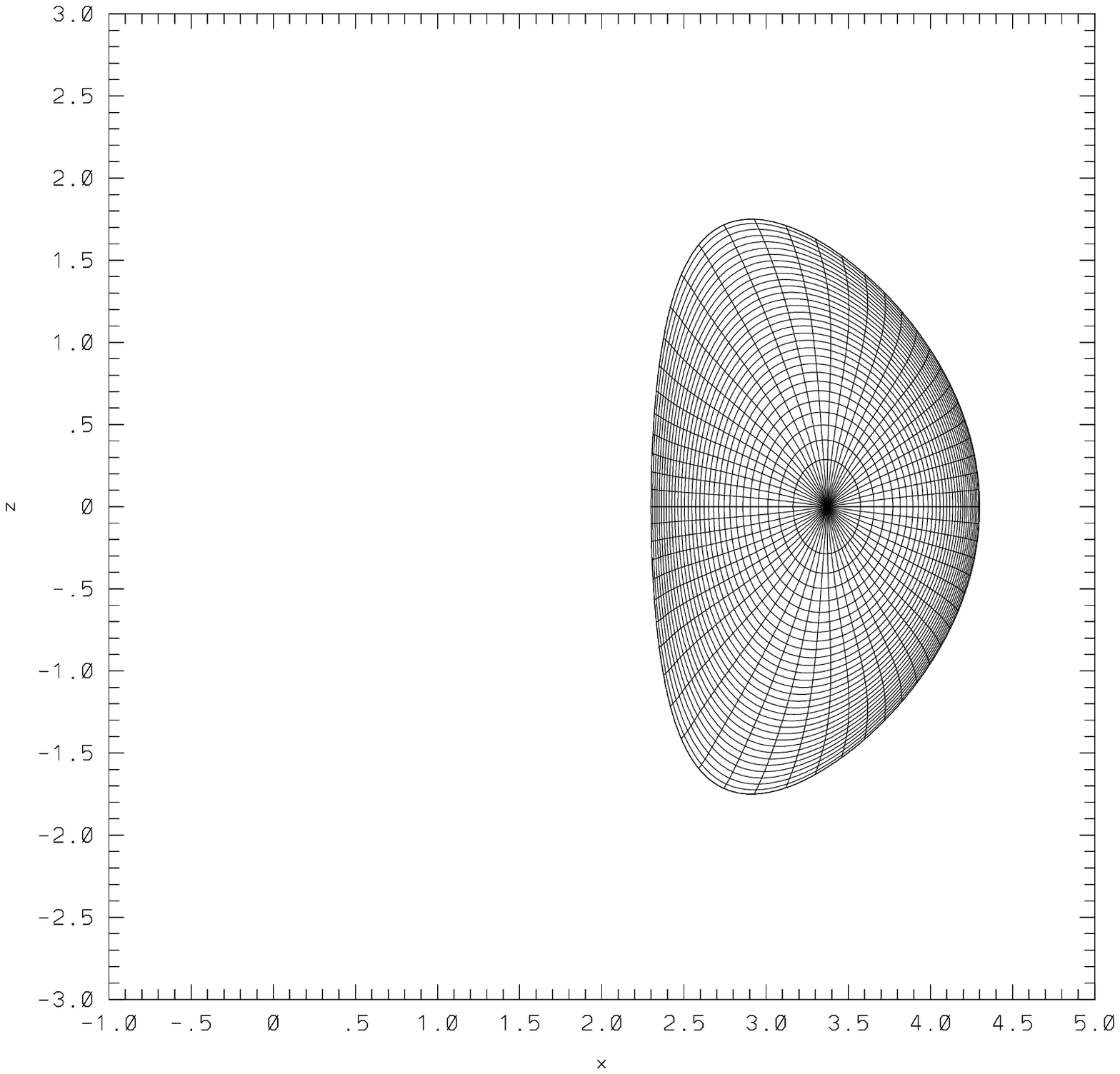}
\end{figure}

\centerline{  Fig.~~ 1}

\newpage

\begin{figure}[htp]
\centering
\includegraphics[width=120mm]{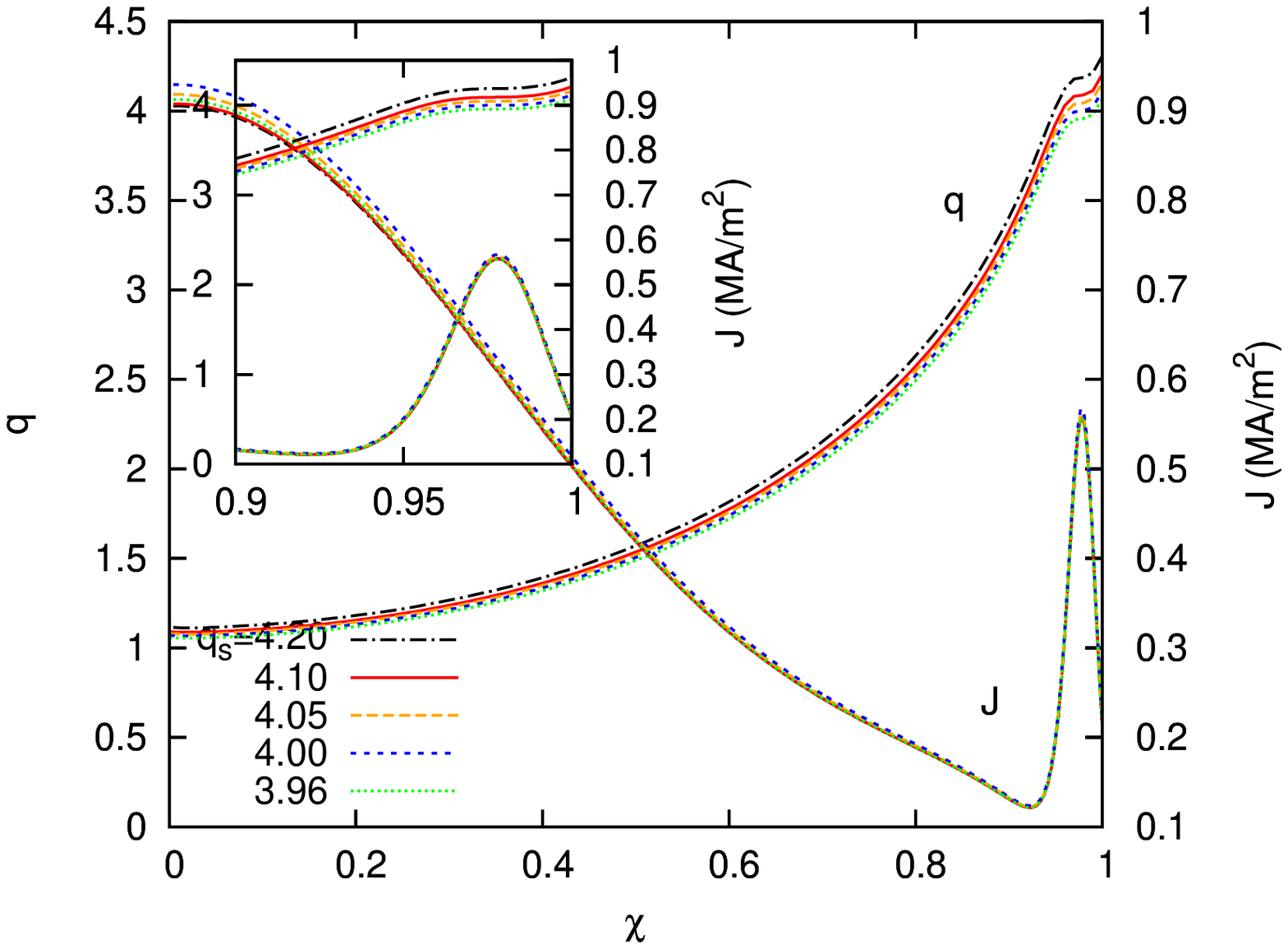}
\end{figure}

\centerline{  Fig.~~ 2}

\newpage

\begin{figure}[htp]
\centering
\includegraphics[width=120mm]{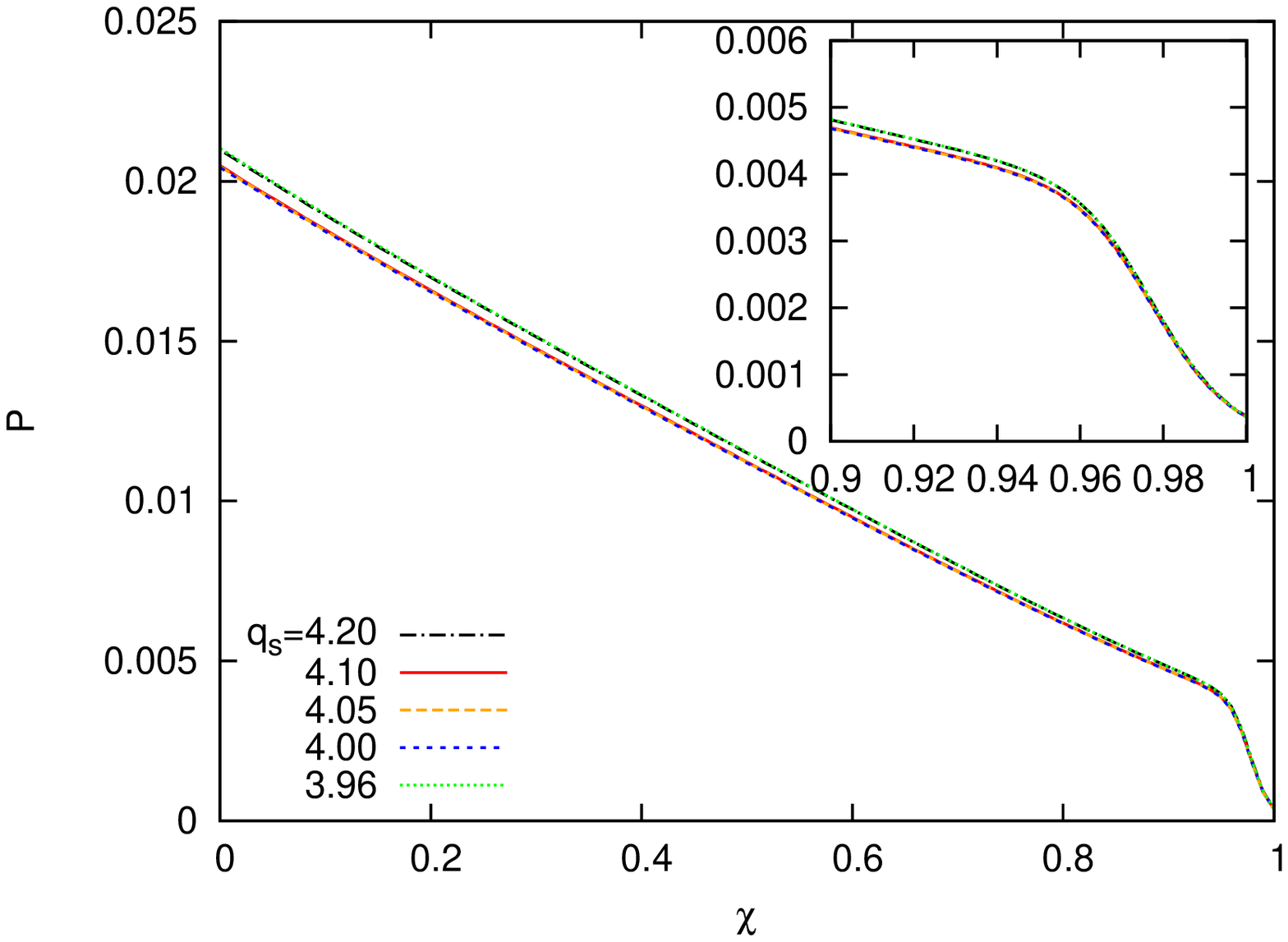}
\end{figure}
\centerline{  Fig.~~ 3}

\newpage

\begin{figure}[htp]
\centering
\includegraphics[width=120mm]{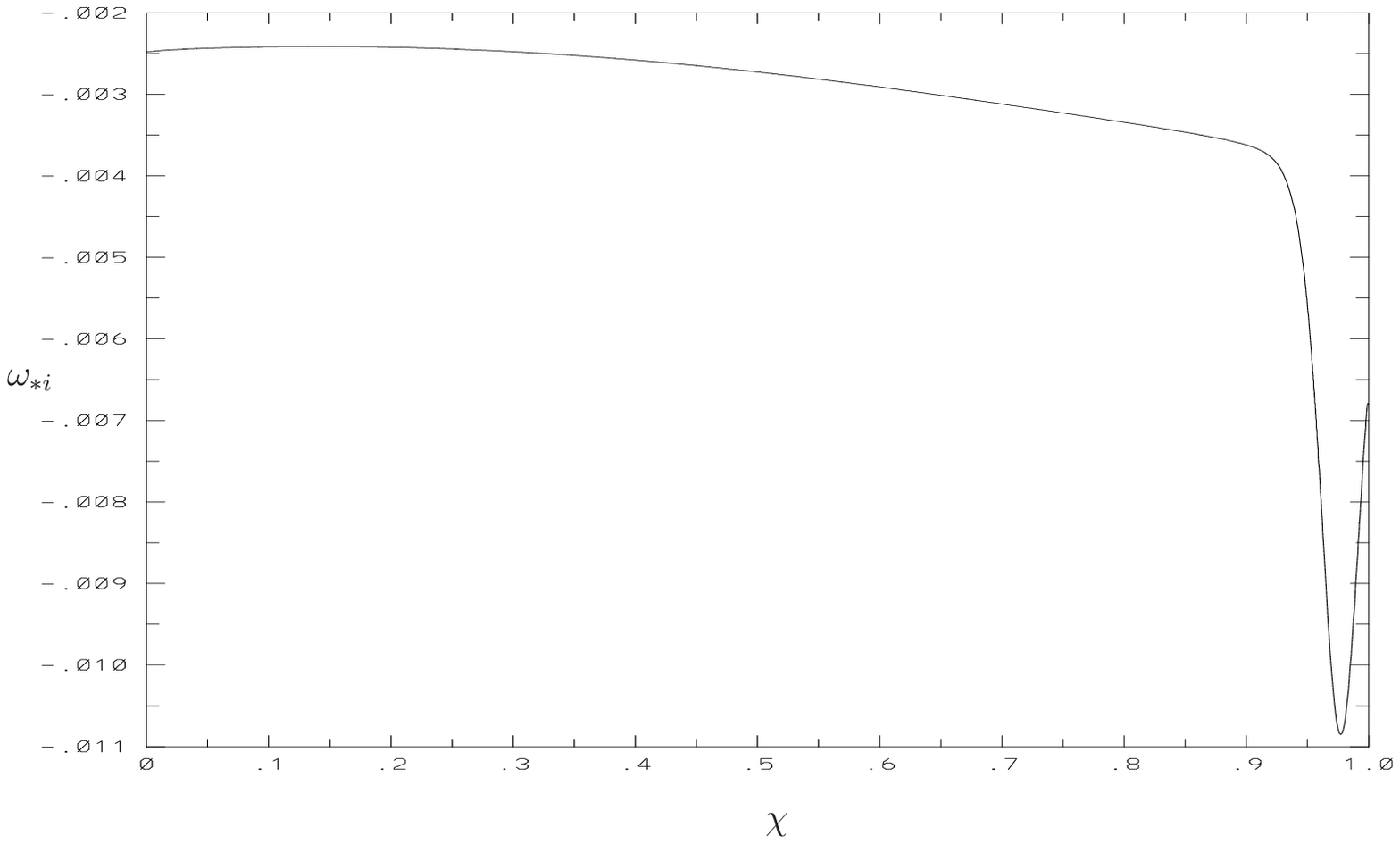}
\end{figure}

\centerline{  Fig.~~ 4}

\newpage

\begin{figure}[htp]
\centering
\includegraphics[width=120mm, angle=-90]{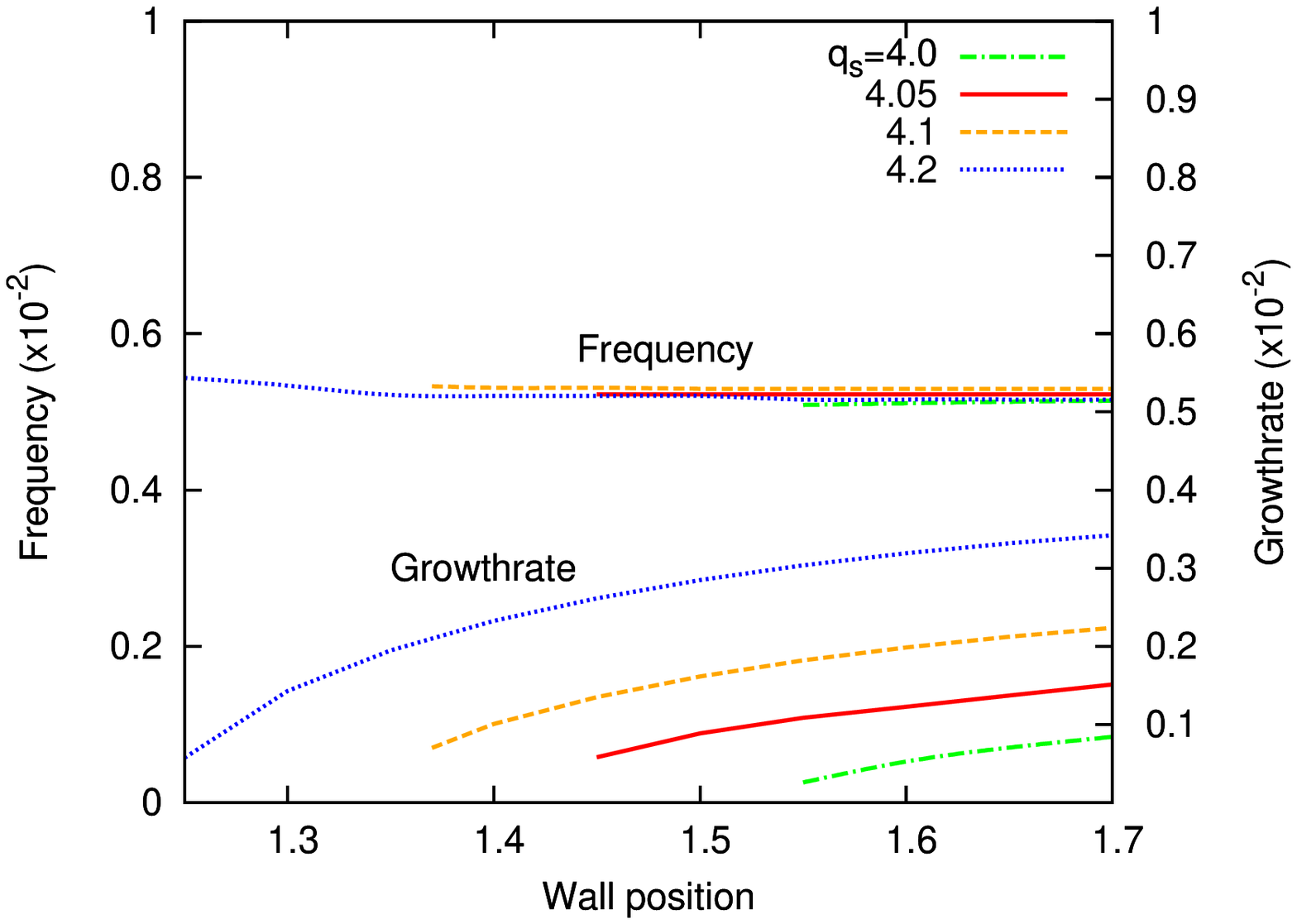}
\end{figure}

\centerline{  Fig.~~ 5}

\newpage

\begin{figure}[htp]
\centering
\includegraphics[width=120mm, angle=-90]{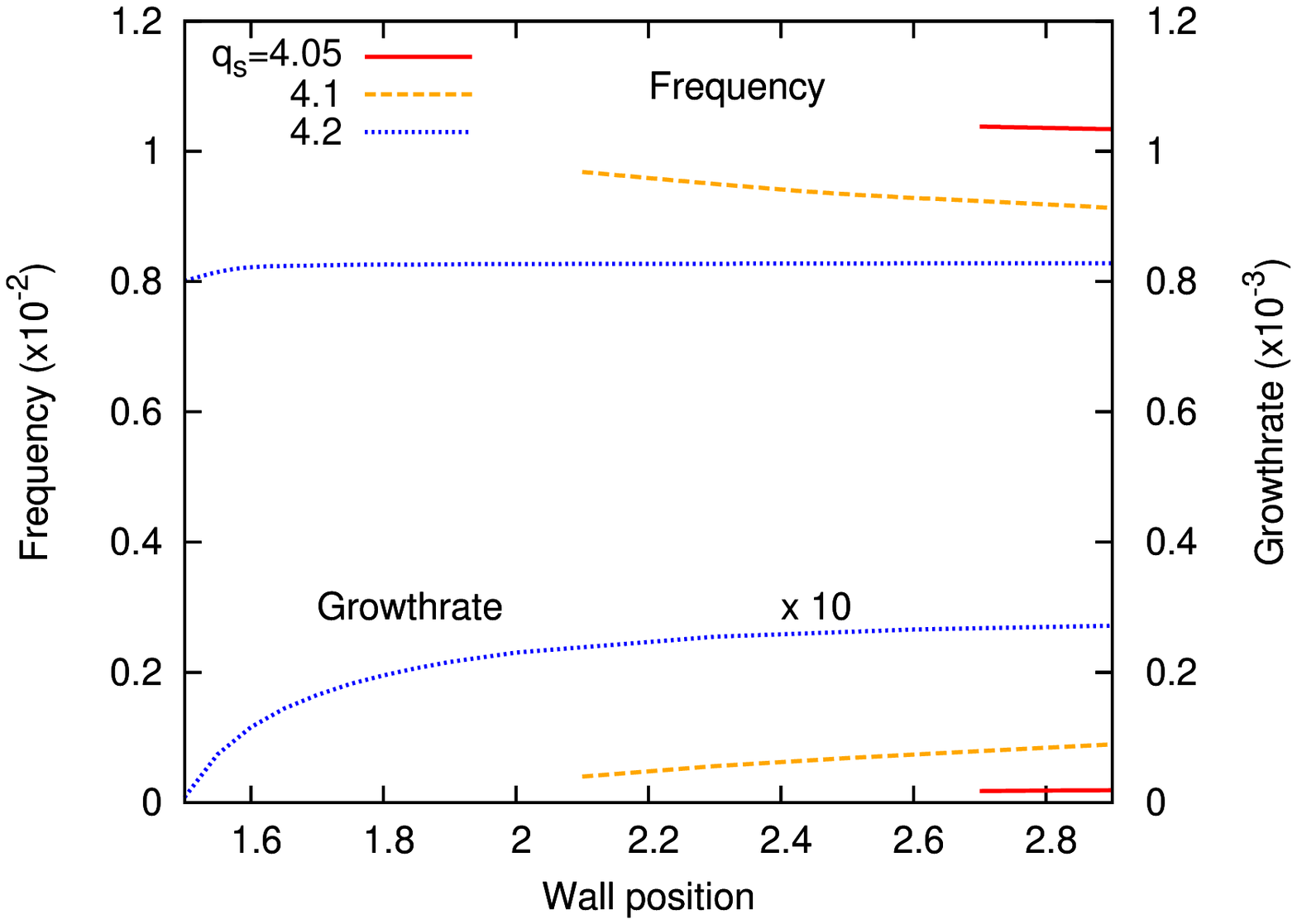}
\end{figure}

\centerline{  Fig.~~ 6}

\newpage

\begin{figure}[htp]
\centering
\includegraphics[width=120mm]{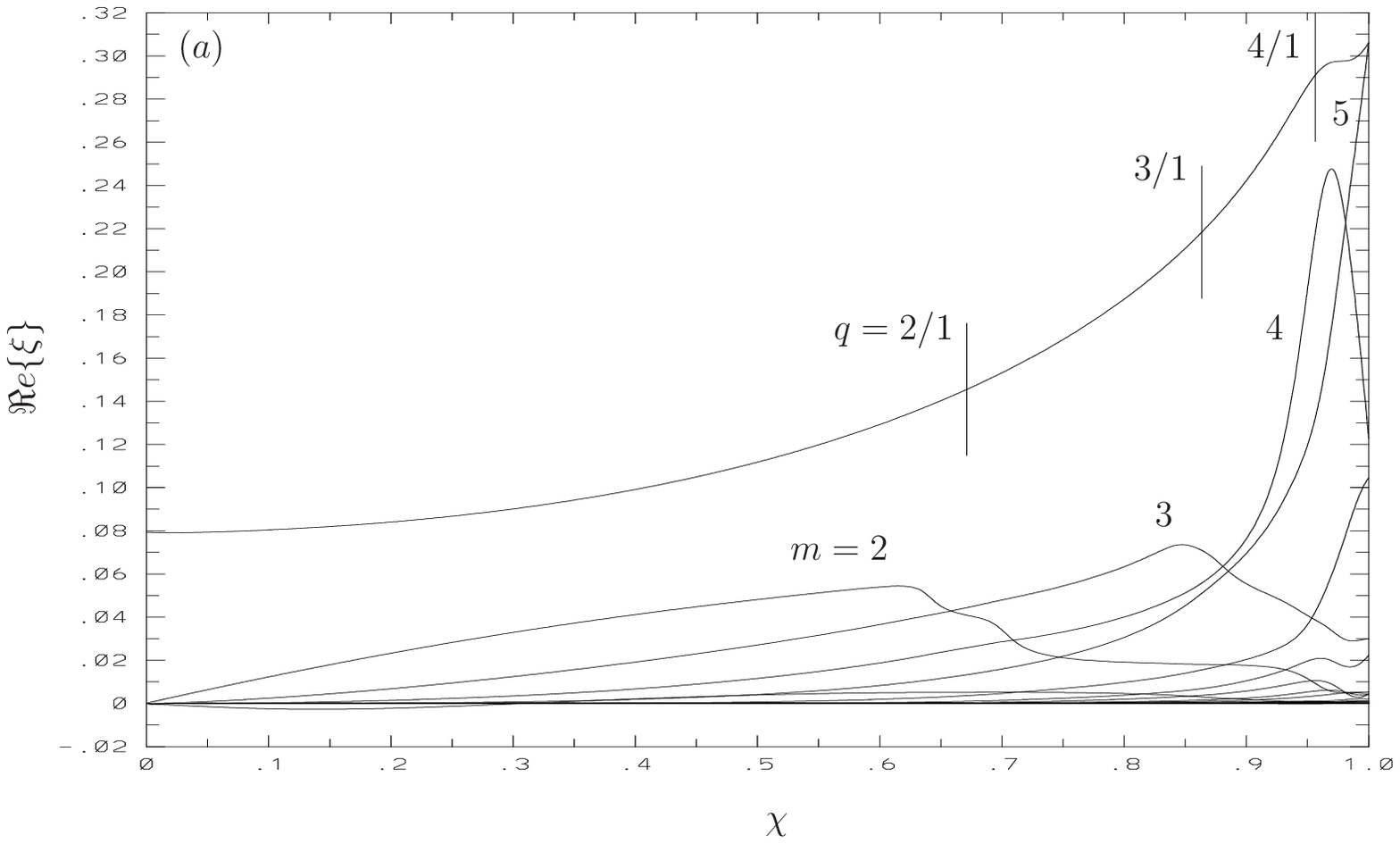}
\end{figure}

\centerline{  Fig.~~ 7a}

\newpage

\begin{figure}[htp]
\centering
\includegraphics[width=120mm]{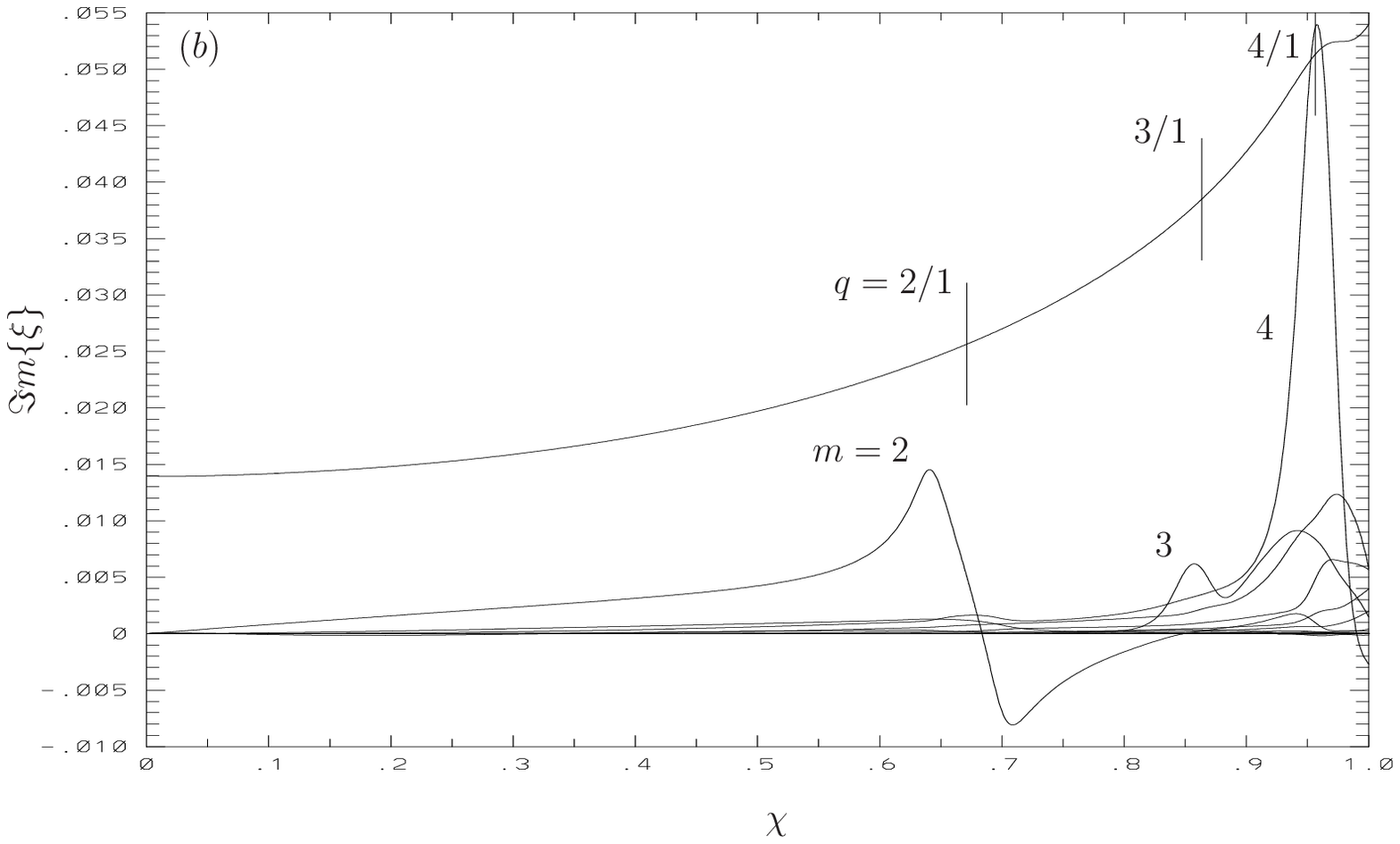}
\end{figure}

\centerline{  Fig.~~ 7b}

\newpage

\begin{figure}[htp]
\centering
\includegraphics[width=120mm]{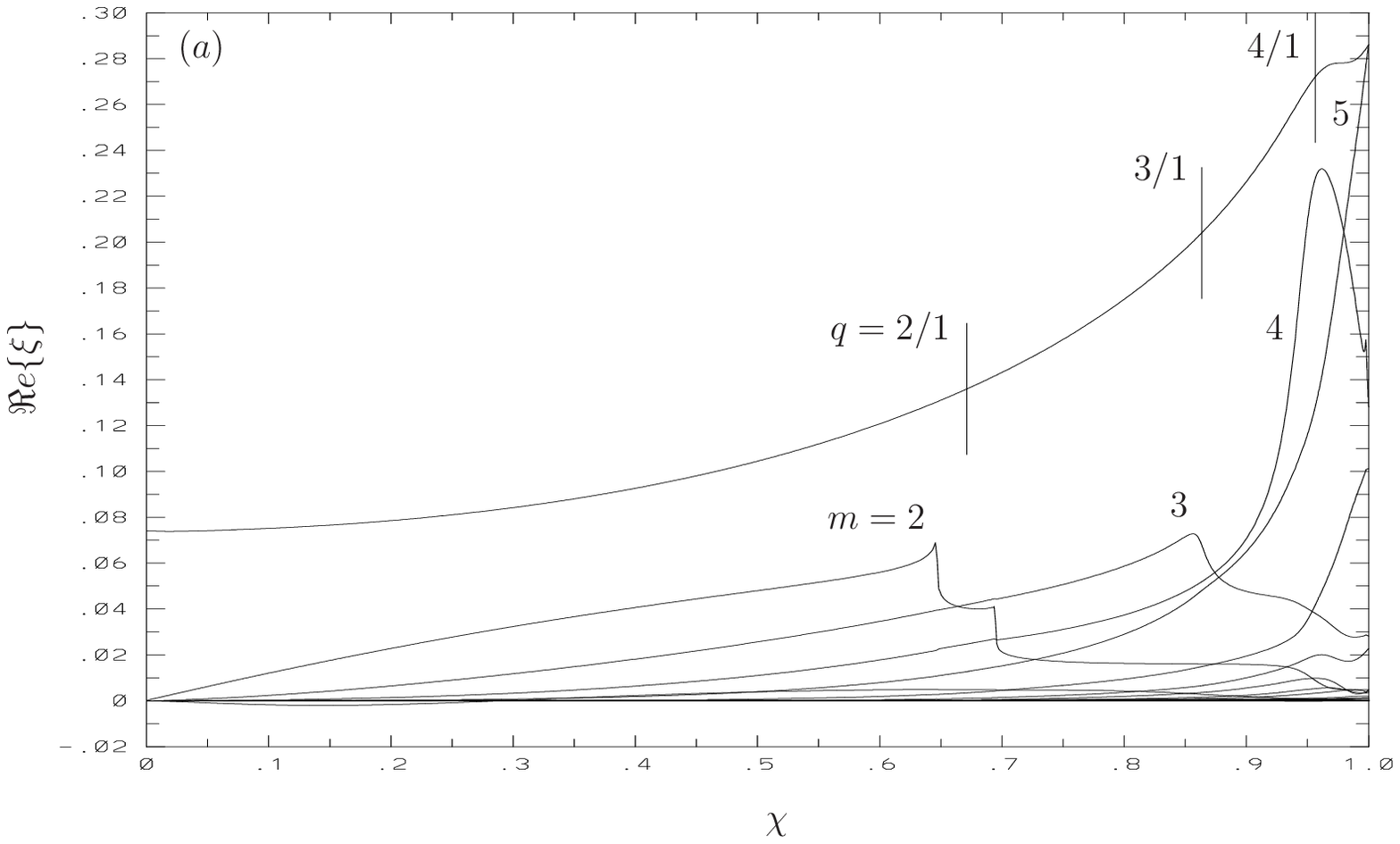}
\end{figure}

\centerline{  Fig.~~ 8a}

\newpage

\begin{figure}[htp]
\centering
\includegraphics[width=120mm]{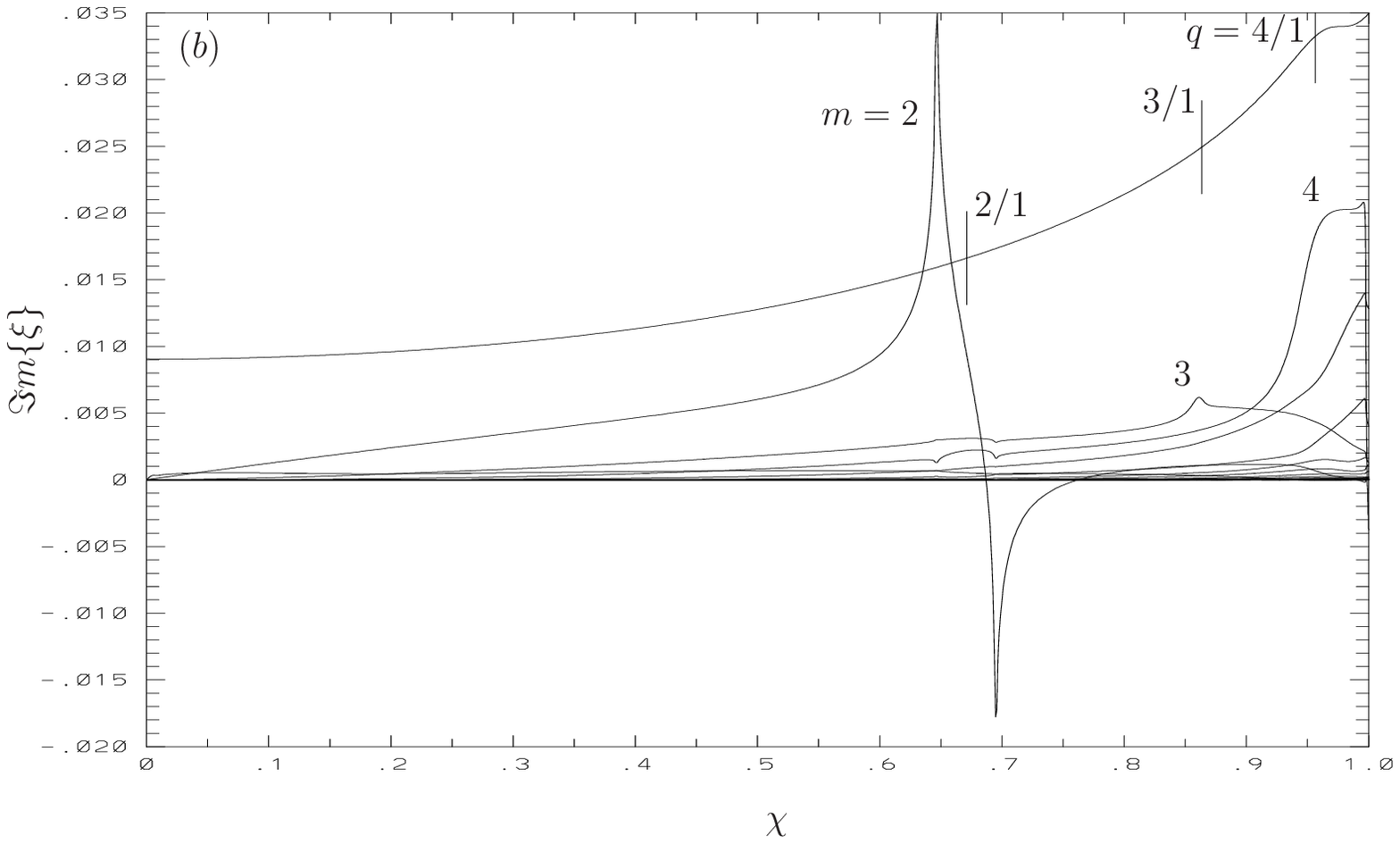}
\end{figure}

\centerline{  Fig.~~ 8b}

\newpage

\begin{figure}[htp]
\centering
\includegraphics[width=120mm, angle=-90]{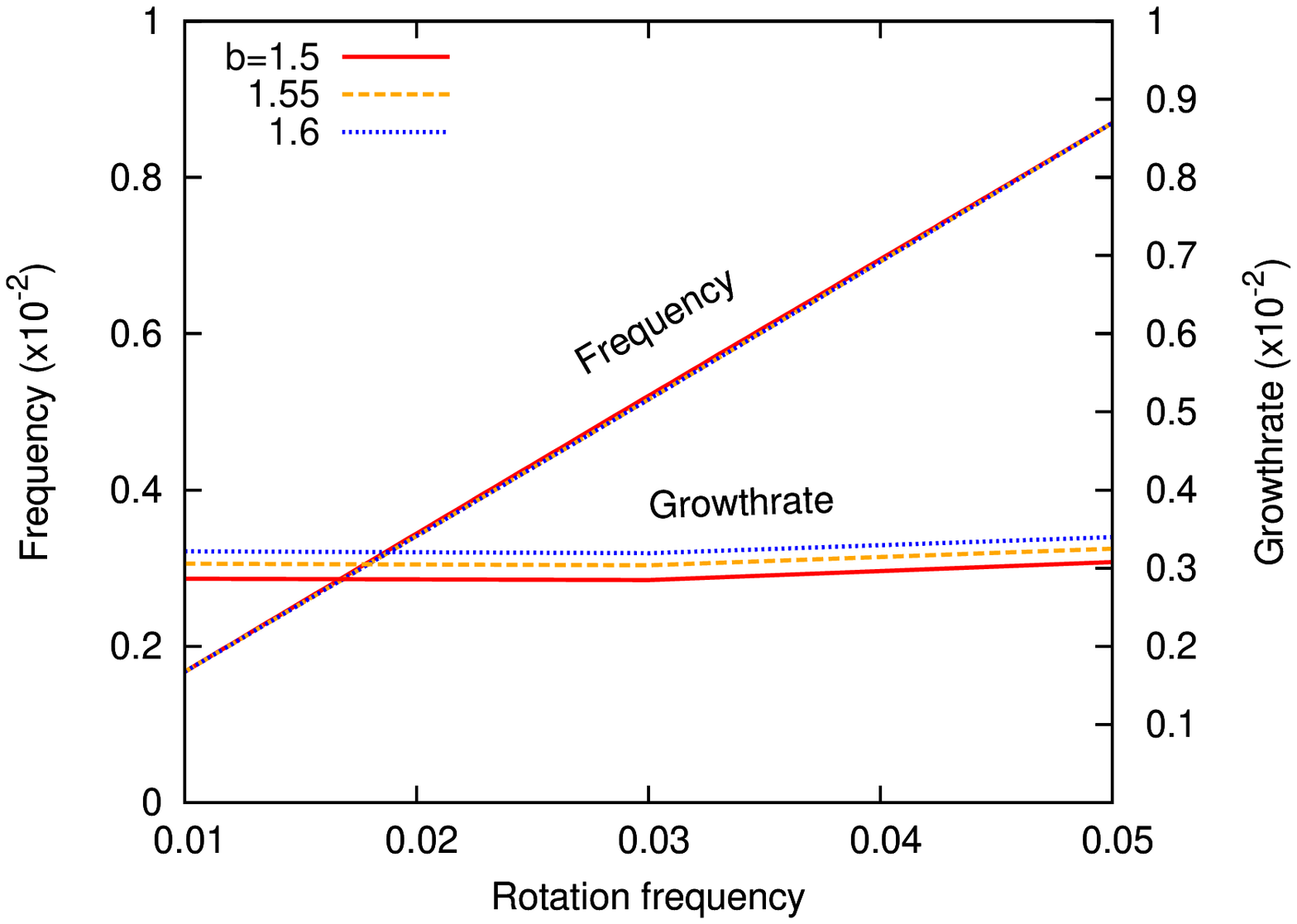}
\end{figure}

\centerline{  Fig.~~ 9}

\newpage

\begin{figure}[htp]
\centering
\includegraphics[width=120mm, angle=-90]{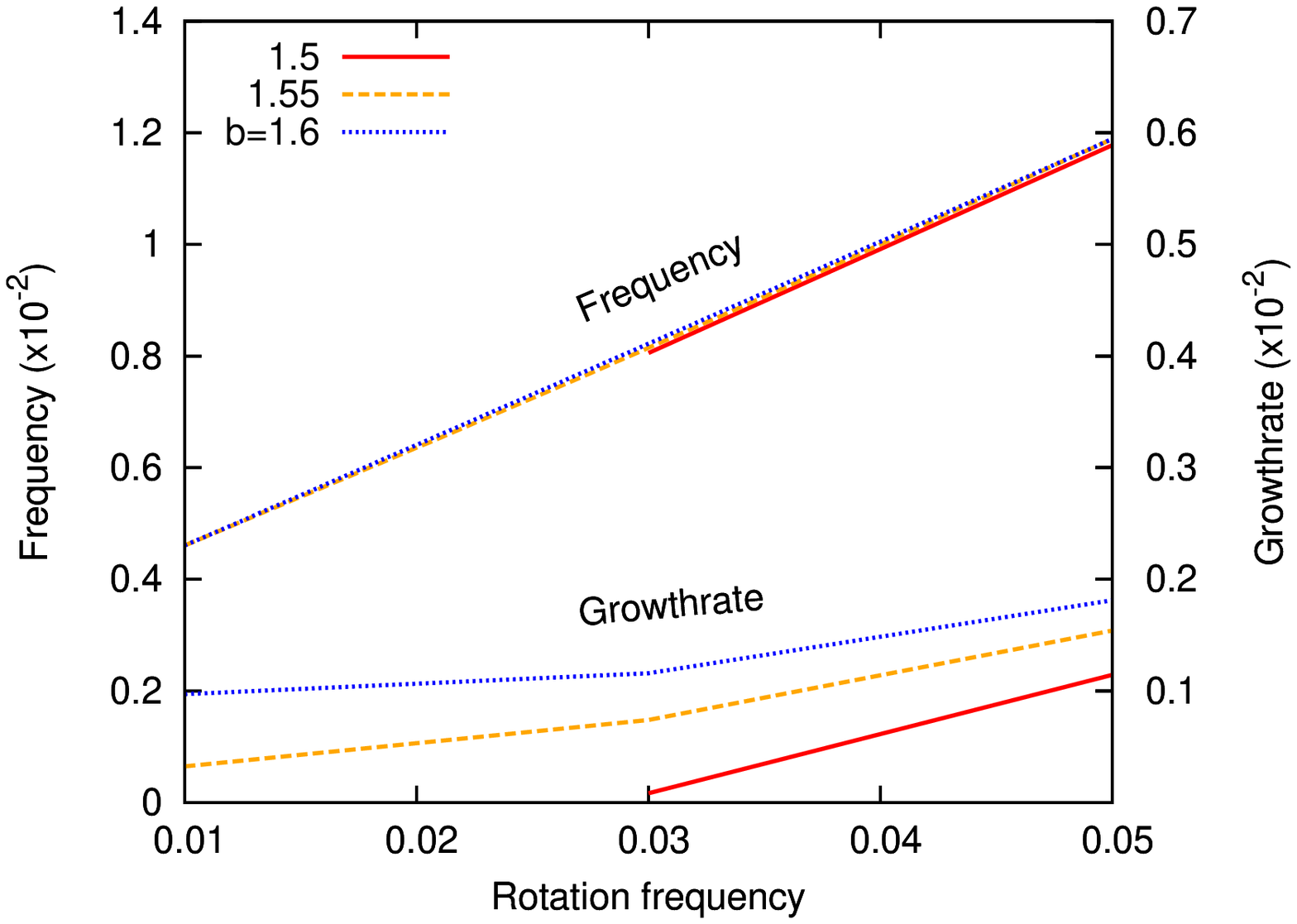}
\end{figure}

\centerline{  Fig.~~ 10}

\newpage

\begin{figure}[htp]
\centering
\includegraphics[width=120mm, angle=-90]{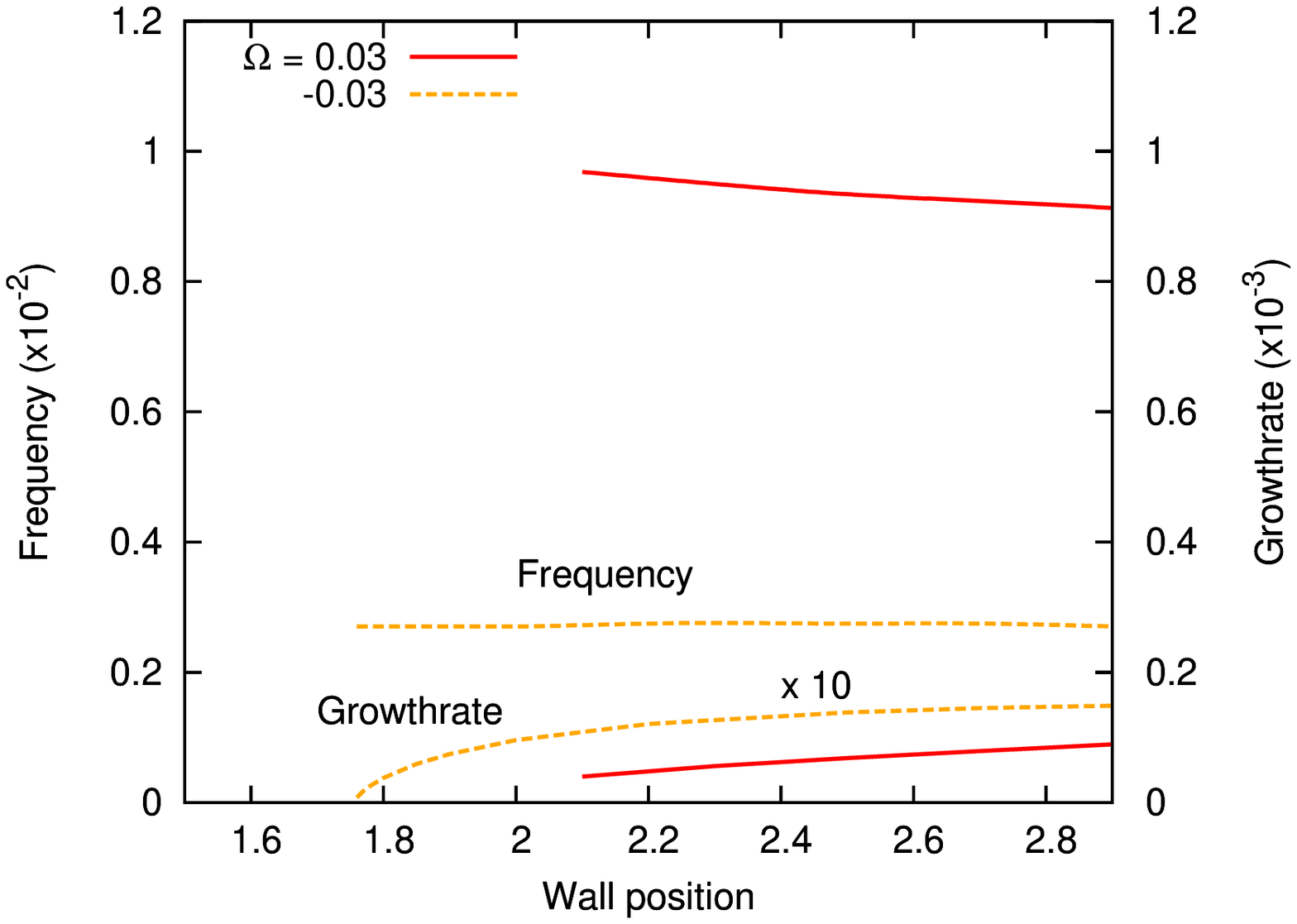}
\end{figure}

\centerline{  Fig.~~ 11}

\newpage

\begin{figure}[htp]
\centering
\includegraphics[width=120mm, angle=-90]{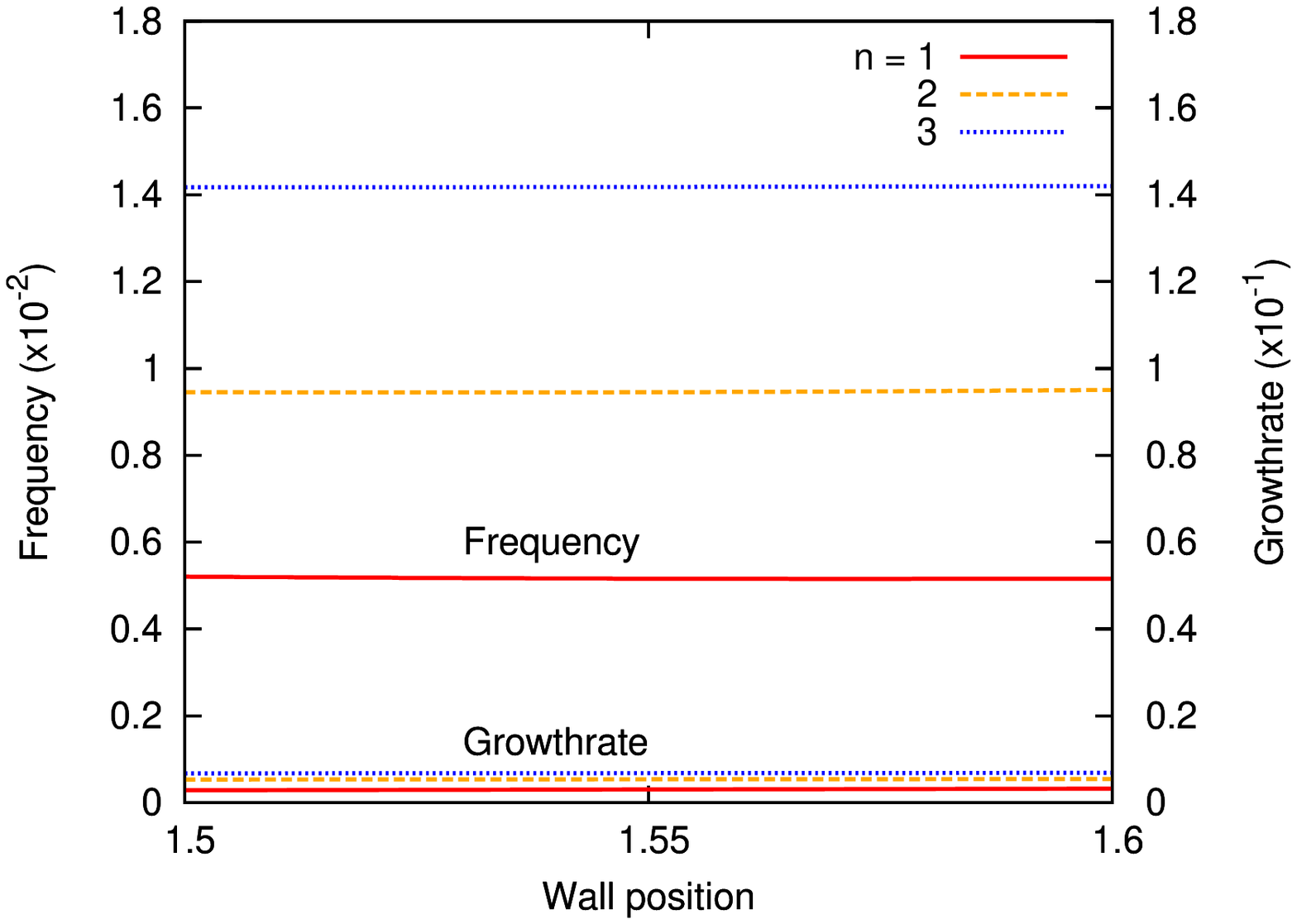}
\end{figure}

\centerline{  Fig.~~ 12}

\newpage

\begin{figure}[htp]
\centering
\includegraphics[width=120mm, angle=-90]{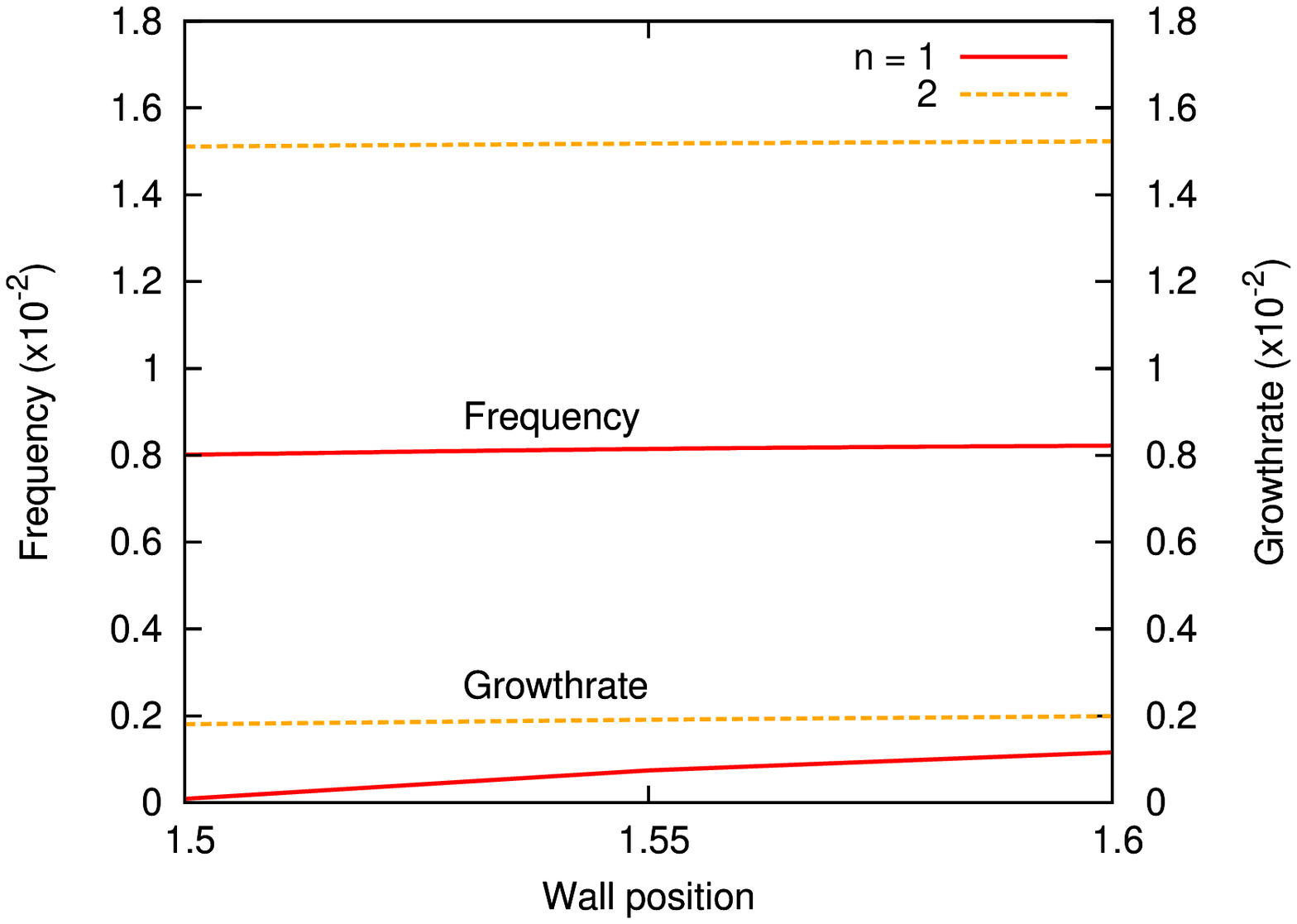}
\end{figure}

\centerline{  Fig.~~ 13}

\end{document}